\documentclass{camera}
\usepackage{epsfig}

\begin{document}
\renewcommand{\thefootnote}{\fnsymbol{footnote}}

%
\title{STATUS AND PROSPECTIVES OF HADRONIC CROSS SECTION AT LOW ENERGY$^*$}

%
\author{Graziano Venanzoni}

%
\organization{Universit\`a and Sezione INFN, Pisa, Italy \\
e-mail:Graziano.Venanzoni@pi.infn.it}

\maketitle

%
\begin{abstract}
In this talk I will review the experimental status and the prospects 
for the future of 
$\sigma(e^+ e^- \to hadrons)$ at low energy.
The recent preliminary results obtained by KLOE and BABAR
collaborations using the radiative return will be also discussed.
\end{abstract}
\footnotetext[1]{Invited talk at the XIV Italian Conference on
  High Energy Physics ``IFAE 2002'', Parma, Italy, April 3-5 2002, to appear in the
  proceedings.}

\noindent
\section{Importance of  hadronic cross section  at low energy}
In the last years a renewed interest has arisen on the measurement of 
hadronic cross section at low energy ($\sqrt{s}<10\, GeV$), mostly because
of  the 
precision reached to test the Standard Model at LEP and SLC~\cite{tenchini} 
 and of the new experimental result of the muon anomalous magnetic moment $a_{\mu}$
at Brookhaven~\cite{Brown:2001mg,Bennett:2002jb}.
The experimental 
accuracy reached so far asks for a precise determination of the theoretical
estimate of both  $\alpha_{QED} (M^2_{Z})$ and $a_{\mu}$, whose main
error comes from the non-perturbative computation of low energy hadronic
contributions to the vacuum polarization, which can be computed using
$e^+ e^-$ data at low energy.
%

The  amazing precision (of $0.7\, ppm$) reached on 
 $a_{\mu}$ at BNL has required  a 
careful re-evaluation of the theoretical 
estimate: as an example, a large part of the 2.6 $\sigma$ deviation
between the 
measurement of   $a_{\mu}$ published last year~\cite{Brown:2001mg}  and 
the Standard Model prediction was caused by a wrong
sign in the calculation of light-by-light scattering 
contributions~\cite{pallante}.

After this sign correction, however, 
the situation for $a_{\mu}^{had}$ is still unclear:
using only $e^+ e^-$ data 
 a $3\,\sigma$ discrepancy between the BNL measured value an the theoretical 
prediction is still 
present~\cite{fj,Davier:2002dy,Hagiwara:2002ma}, 
which becomes 1.6 $\sigma$ if hadronic decays of
$\tau$ are also used~\cite{Davier:2002dy}. 
\section{Status of hadronic cross section  at low energy}
\begin{figure}[h]
\begin{tabular}{cc}
\epsfig{file=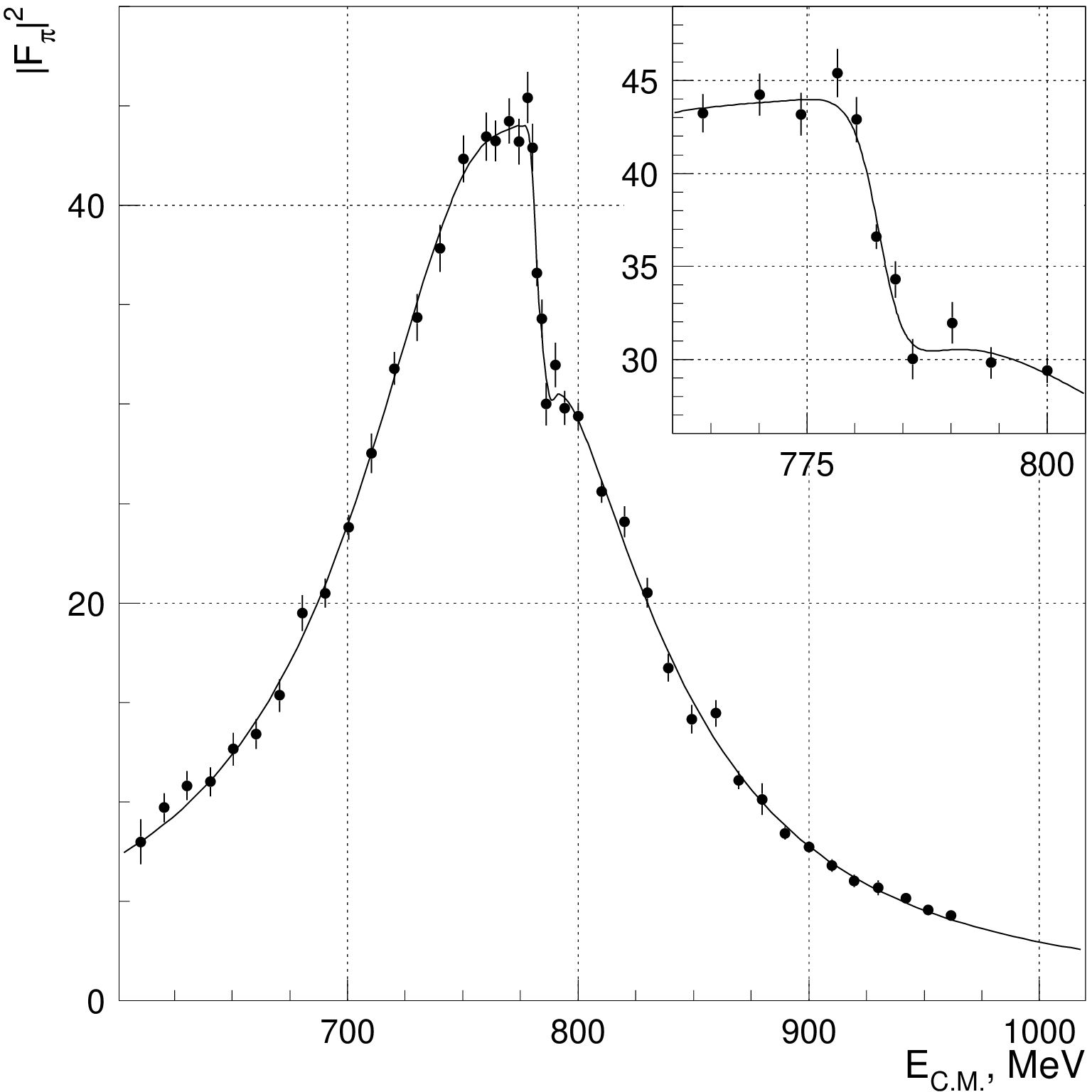,width=6cm,height=6cm} &
\epsfig{file=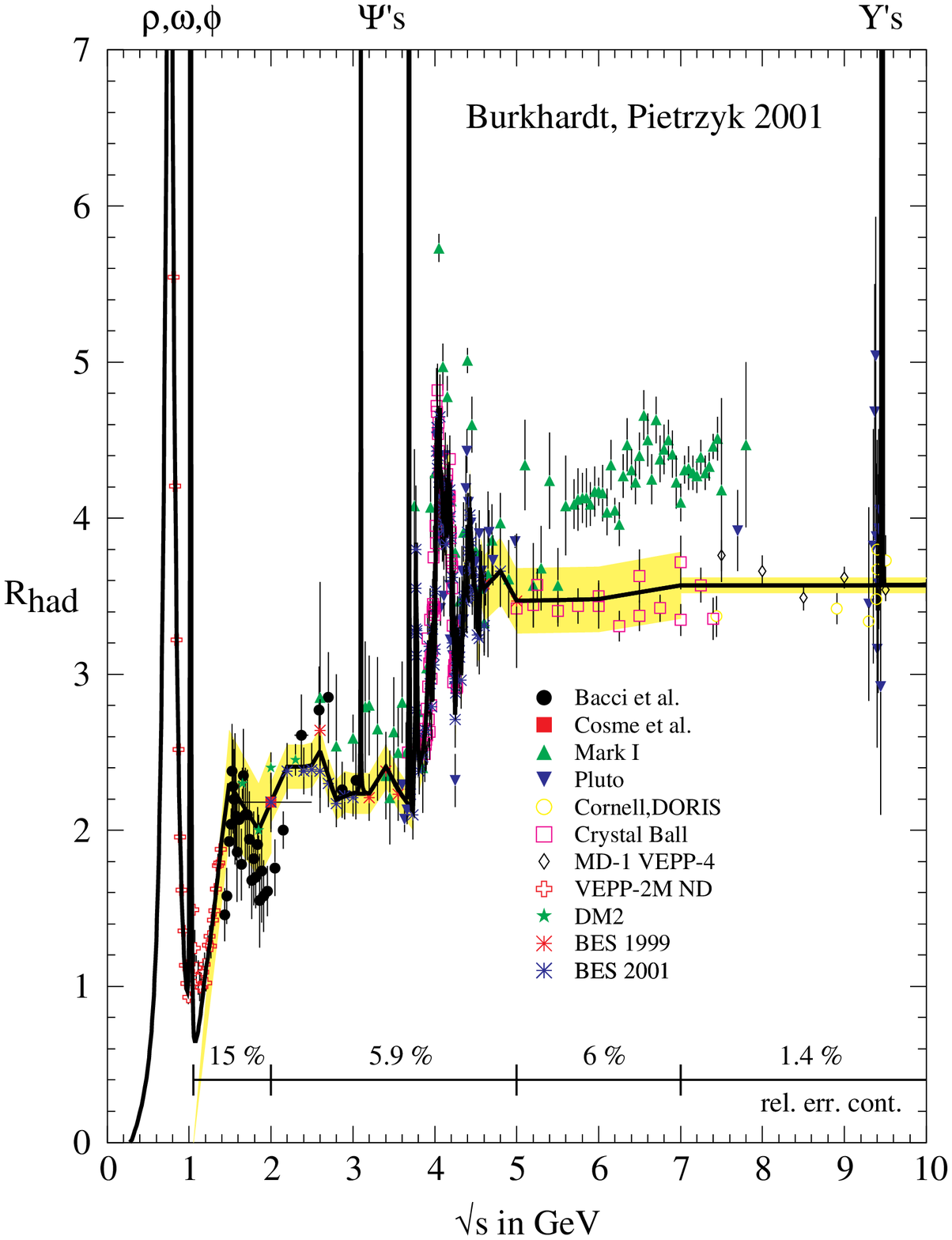,width=6cm,height=6cm} \\
\end{tabular}
\caption{{\it Left}: measurement of the pion form factor from CMD-2. {\it
    Right}: an updated compilation of R-measurement from H. Burkhardt and B. 
Pietrzyk~\cite{Burkhardt:2001xp}.}
\label{fig1}
\end{figure}

Hadronic $e^+ e^-$ annihilation cross sections 
 have been measured
by many laboratories in the last 20 years~\cite{Venanzoni:2001jh}. 
Usually the cross section for individual channel 
is measured for energies below 2 GeV, 
while above that, the hadronic final states are treated inclusively. 
Fig.~\ref{fig1}({\it right}) shows
an up-date compilation of these data done by Burkhardt and 
Pietrzyk~\cite{Burkhardt:2001xp}.
The main improvements come in the region below 5 GeV, in particular
between $2-5\,GeV$ where the BESII coll. has reduced the error 
to $\sim 7\%$~\cite{Bai:2001ct} (before was $\sim 15\%$), 
and in the region below 1 GeV, where the CMD-2 coll. has measured
the pion form factor with 
a  systematical error of $0.6\%$  in the energy range from 0.61 to 
0.96 GeV~\cite{Akhmetshin:2001ig}
(see Fig.~\ref{fig1}({\it left})). Both these new results have  a 
significant impact on 
the updated calculation of $\alpha_{QED} (M^2_{Z})$~\cite{gambino} 
and $a_{\mu}$~\cite{pallante}, and will be discussed below.
While the data between 2-5 GeV are now closer to perturbative QCD,
the error in the 1-2 GeV region is still  15\%:
a reduction of  this error to few {\it percent} will be very important
both for $\alpha_{QED} (M^2_{Z})$ and $a_{\mu}$ calculations.
\subsubsection*{\bf Measurement of pion form factor at 0.6\% from CMD-2}
The CMD-2 collaboration has recently published a new measurement of 
 $\sigma(e^+ e^- \to \pi^+\pi^-)$ at  the VEPP-2M $e^+\,e^-$ collider,
with a 0.6\% systematic uncertainty in the center-of-mass energy range 
from 0.61 to 0.96 GeV~\cite{Akhmetshin:2001ig}. Such an accuracy has 
been mainly possible thanks to  
a careful evaluation of radiative corrections and  to the control of 
different sources of systematic errors.
In  particular: (i) the error coming from the
 energy beam is reduced  by the resonance depolarisation
technique; (ii) the systematic effects on the efficiency evalutation 
(trigger, tracking) are minimized by normalizing $e^+ e^- \to \pi^+\pi^-$ 
events to collinear events (mainly $e^+ e^- \to e^+ e^-$).
Differently from the previous publication, 
the cross section is corrected for leptonic and hadronic 
vacuum polarization and for photon radiation by the emitted pions (FSR).
In order to make meaningful comparison between different data it is   
very important to have a clear definiton of the corrections used.

\subsubsection*{Recent results from BESII at BEPC}
The BESII collaboration has recently published a new measurement of R 
(ratio of hadron to muon production) in
the region 2-5 GeV, based on 85 points taken between February and June 99,
 with an average precision of 6.6\%, a factor 2 better of the previous
results~\cite{Bai:2001ct}. R was determined inclusively, from the number of observed hadronic
events, $N_{had}^{obs}$:
\[
R=\frac{N_{had}^{obs}-N_{bckg}-\sum_l N_{ll}-N_{\gamma\gamma}}
{\sigma^0_{\mu \mu}\cdot L \cdot \epsilon_{had}\cdot (1+\delta)}
\]
where $N_{bckg}$ is the number of beam-associated background events; 
$\sum_l N_{ll}$  and $N_{\gamma\gamma}$ are respectively 
the background coming from misidentified events in one and two photons
processes; $L$ is the integrated luminosity; $\delta$ is the radiative
correction and $\epsilon_{had}$ is the overall detector efficiency.
In order to keep the error to $\sim 7\%$ a big effort was done 
on: (a) Monte Carlo simulation to better understand detector efficiency;
(b) estimation of $N_{bckg}$ by means of separated beam and single beam
  operation; (c) radiative correction by comparing different schemes.

\section{First results with radiative return}
Starting from the papers of V. N. Baier and V. A. Khoze~\cite{bk}, 
where 
the mechanism of returning to a resonant region was
for the first time introduced, in the last years
big efforts from theoretical and experimental side 
have been dedicated  to the measurement  of  hadronic cross sections
at meson  factories using the  radiative return,
 {\it i.e.} studying the process
$e^{+} e^{-} \rightarrow hadrons+\gamma$.
 The emission of a photon in the initial state (ISR) lowers 
the interaction energy and 
 makes possible to produce the hadronic system with an invariant 
mass varying from the meson ($\phi$, $\Upsilon(4S)$)
mass down to the production threshold. 
High luminosity of the machine compensates for the 
reduced cross section.

The method~\cite{Binner:1999bt,spagnolo} represents 
an alternative and competitive 
approach to the conventional  energy scan:  
it has the advantage of the same normalization for each energy 
point, even if it needs a very solid theoretical understanding 
of ISR (including radiative corrections), as well as a
precise determination of the angle and energy of the emitted photon 
and the full control of background events, expecially for 
events with the photon  emitted in the final state (FSR).
Radiative corrections
have been calculated up to NLO by different 
groups~\cite{Rodrigo:2002rz,Khoze:2002ix,Hoefer:2001mx}, for the exclusive
final hadronic state $\pi^{+}\pi^{-}\gamma$ (yielding event generator 
EVA~\cite{Binner:1999bt}, PHOKHARA~\cite{Rodrigo:2001kf}) and $4\pi+\gamma$~\cite{Czyz:2000wh}. 

Both KLOE experiment at the Frascati $\phi$-factory DA$\Phi$NE and 
subsequently BABAR at SLAC
have already presented promising results using this method.

The KLOE analysis is focusing on $e^{+}e^{-} \rightarrow \pi^{+}\pi^{-}\gamma$,
from which  $\sigma(e^{+}e^{-} \rightarrow \pi^{+}\pi^{-})$ 
can be extracted. The momenta of the two pions 
are measured with the Drift Chamber ($\sigma_{p_T}/p_T\leq 0.4\%$), while the photon is not tagged: 
its energy and direction is retrieved by a 
kinematical constraint on the missing momentum of the event. 
This allows to have a large statistics
and a reduced background 
(in particular for FSR, which is below 1\%). KLOE has already collected
$\sim 500\, pb^{-1}$: the first results, based on less than 1/6 
of the total statistics show an error for single point of 2\% above 0.4 
$GeV^2$~\cite{Venanzoni:2002jb}.
 
Due to the different experimental technique, 
a comparison between KLOE and CMD-2 will be very important,  
particularly after the recent discrepancy 
 between  $e^+e^-$ and isospin breaking-corrected 
$\tau$ data~\cite{Davier:2002dy}.
It must be however kept in mind that a precise comparison between 
these experiments
will be significant only if the same corrections are applied to the data:
the inclusion of  FSR corrections means 
in particular the evaluation of the KLOE efficiency for two 
photon final states (one from initial and one from the final state).

BABAR analysis can benefit from the large collected statistics 
($\sim 90\,fb^{-1}$)
which corresponds to $\sim 3.6$ million fiducial events with a hadronic system 
invariant mass less than $7\,GeV$~\cite{Berger:2002mg}.
Using the radiative return, it is  possible to perform both an inclusive analysis, 
which would rely only 
on the identification of the tagged initial-state radiation (ISR) photon,
 as well as  
an exclusive analysis, in which each possible hadronic decay channel
is analysed separately. BABAR has already presented some preliminary results 
 both for the $2\pi$ and $4\pi$ 
final states~\cite{babar}.

BABAR results, which  are expected to cover many different channels, 
will be in particular important in the region 
between 1.4 and  2 GeV, where the discrepancy between different experiments
is at the level of 15-20\%.

\section{Prospects for the future}
In the next years the measurament of hadronic cross section 
at low energy should benefit from additional
data both from the current experiments as well as from 
new projects:
(1) VEPP-2000, which is 
a new collider proposed at Novosibirsk with $\sqrt{s}$ up to 2 GeV and
  expected luminosity of $10^{31}-10^{32} cm^{-2} sec^{-1}$; 
(2) CLEO-C, a  modification of CESR for a high
 luminosity machine in 3-5 GeV region~\footnote{  
{\tt http://www.lns.cornell.edu/public/CLEO/CLEO-C/index.html}},
and (3) BESIII/BEPCII, an upgrade for 
BEPC collider and BES detector.

Tab.~\ref{tab2} shows the prospects for R 
in the energy range below 5 GeV.

\begin{table}
\begin{center}
\noindent
\begin{tabular}[h]{|c|c|c|c|}
\hline\hline
Energy  & Current & Expected & Expected \\
Range (GeV)       &  Error   & Results  & error \\
\hline
$0.3-0.6$  ~   &1-2\%  & CMD-2, VEPP-2000, KLOE, B-factories & 1\%? \\
\hline
    $0.6-1$ & 0.6\% & KLOE, B-factories, VEPP-2000 & $<0.6\%?$ \\  
\hline
$1 - 1.4$   &5-10\%  & CMD-2, SND, B-factories, VEPP-2000   & $<5\%?$ \\
\hline
$1.4 - 2$ & 15-20\% & B-factories, VEPP-2000  & 5\%? \\
\hline
$2 - 5$  & 7\% &  BEPCII, CLEO-C, B-factories  &3\% \\
\hline
\end{tabular}
\caption{Prospects on R measurement at low energy.}
\label{tab2}
\end{center}
\end{table}

All the future results
will contribute to determine an new exciting era
for the hadronic cross section measurement.

\newpage
\end{document}